# Effects of the interaction between the magnetic moments on the energy levels of a pionic atom.

Voicu Dolocan

*Faculty of Physics, University of Bucharest, Bucharest, Romania*

We present the calculation of the binding energy of 2p pion for various central charges Z. We have taken into account the interaction between the magnetic moment of the nucleus and the orbital magnetic moment of the pion. The obtained results show that the interaction between the magnetic moments decreases the absolute value of the binding energy. This decrease is larger when the two magnetic moments add one another. Also, the binding energy increases with Z increasing.

## 1. Introduction

It has been of interest for physicists to study bound states of charged pions to nuclei, "pionic atoms". The pion has the mass $mc^2$ =139.577 MeV and spin 0. It obeys the Klein-Gordon equation

$$[(E-V(r))^2 - m^2 c^4 + \hbar^2 c^2 \nabla^2]\psi(\mathbf{r}) = 0 \qquad (1)$$

The time independent field equation (1), in the static Coulomb field, $V(r) = -Ze^2/r$, for the radial wave function $\Psi = R(r) Y_l^m$ has the same form as the non-relativistic Schrödinger equation

$$[\frac{\hbar^2 c^2}{2E}(\frac{-d^2}{dr^2} - \frac{2}{r}\frac{d}{dr} + \frac{l(l+1) - Z^2\alpha^2}{r^2}) - \frac{Ze^2}{r}]R = \frac{E^2 - (mc^2)^2}{2E}R \qquad (2)$$

By comparing to the Schrödinger hydrogen-like equation

$$[\frac{\hbar^2}{2\mu}(\frac{-d^2}{dr^2} - \frac{2}{r}\frac{d}{dr} + \frac{\lambda(\lambda+1)}{r^2}) - \frac{Ze^2}{r}]R = \epsilon R \qquad (3)$$

we find

$$\mu = \frac{E}{c^2}$$

$$\lambda = \sqrt{(l+\frac{1}{2})^2 - Z^2\alpha^2} - \frac{1}{2}$$

$$\epsilon = \frac{E^2 - (mc^2)^2}{2E} \qquad (4)$$

Here $\lambda$ is not an integer. As a result the bound state eigenvalues are given by

$$\epsilon = \frac{-1}{2}\frac{Z^2\alpha^2\mu c^2}{\nu^2}$$

where $\nu$ is the principal quantum number and takes values $\nu = \lambda+1, \lambda+2, \ldots$ Next,

$$\frac{E^2-(mc^2)^2}{2E}=\frac{-1}{2}\frac{Z^2\alpha^2 E}{\nu^2} \qquad (5)$$

$\alpha = e^2/4\pi\varepsilon_o\hbar c = 1/137.03602$ is the fine structure constant. Solving for $E$ one obtains

$$E=\frac{mc^2}{\sqrt{1+Z^2\alpha^2/\nu^2}}$$

Expanding the above eigenvalue in a series of powers of Z yields

$$E=mc^2(1-\frac{1}{2}\frac{Z^2\alpha^2}{\nu^2}+\frac{3}{8}\frac{Z^4\alpha^4}{\nu^4}+...)$$

By expanding $\lambda$ up to $O(Z^2\alpha^2)$ we can write

$$\lambda=1-\frac{Z^2\alpha^2}{2l+1}+...$$

and

$$\nu=n-\frac{Z^2\alpha^2}{2l+1}+...$$

Therefore [1],

$$E=mc^2(1-\frac{1}{2}\frac{Z^2\alpha^2}{n^2}-\frac{Z^4\alpha^4}{(2l+1)n^3}+\frac{3}{8}\frac{Z^4\alpha^4}{n^4}+...) \qquad (5)$$

where $n = l +1, l +2, \ldots$ The second term is the term we obtain in non-relativistic Schrödinger equation. The next two terms are the so-called "relativistic correction", obtained by expanding the relativistic kinetic energy

$$\sqrt{p^2c^2+(mc^2)^2}=mc^2+\frac{p^2}{2m}-\frac{1}{8}\frac{(p^2)^2}{8m^3c^2}+... \qquad (6)$$

May be written

$$\boldsymbol{p}^2\psi=2m(\frac{Ze^2}{r}-\frac{1}{2}\frac{Z^2\alpha^2 mc^2}{n^2})$$

and therefore

$$\langle\Psi|\frac{-1}{8}\frac{(\boldsymbol{p}^2)^2}{m^3c^2}\Psi\rangle=\frac{-1}{2mc^2}\langle\Psi|(\frac{Ze^2}{r}-\frac{Z^2\alpha^2 mc^2}{2n^2})^2\psi\rangle \qquad (7)$$

Using

$$\langle\Psi|\frac{1}{r}\Psi\rangle=\frac{1}{n^2 a_o} \qquad \langle\Psi|\frac{1}{r^2}\Psi\rangle=\frac{2}{(2l+1)n^3 a_o^2}$$

with $a_o = 4\pi\varepsilon_o \hbar^2/mZe^2$, one obtains

$$\langle\Psi|\frac{-(\mathbf{p}^2)^2}{8m^3 c^2}\Psi\rangle = mc^2\left(\frac{-Z^4\alpha^4}{(2l+1)n^2}+\frac{3}{8}\frac{Z^4\alpha^4}{n^4}+...\right) \qquad (8)$$

Because the Klein-Gordon field has not spin, there is no spin-orbit coupling. Next, we calculate the energy levels of a pionic atom for $n \geq 2$ and $l \geq 1$, where play an important role the interaction between the magnetic moment of the nucleus and the orbital magnetic moment of the pion. We use a modified Coulomb potential due to the interaction between the magnetic moments as we have proceed to study ferromagnetism [2], high excitation energy levels of helium [3], deuteron energy states[4], Lamb shift in hydrogen atom [5] and Yukawa potential[6].

## 2. Effects of the interaction between the magnetic moments.

The magnetic moment of the nucleus is expressed in terms of the nuclear magneton $\mu_N$ and nuclear spin I in the form

$$\mu = \gamma_N \mu_N I$$

where

$$\mu_N = \frac{e\hbar}{2M_p c}$$

Experimentally values of $\gamma_N$ for some nuclei are presented in Table I [7,8]

Table I

Nuclear constants, I, $\gamma_N$ for some nuclei

| Z  | M, a.u. | I   | $\gamma_N$ |
|----|---------|-----|------------|
| 10 | 20.183  | ½   | -1.88542   |
| 20 | 40.08   | 3   | 1.6022     |
| 30 | 65.37   | 5/2 | 0.76902    |
| 40 | 91.22   | 5/2 | -1.30362   |
| 50 | 118.69  | 2   | 0.042      |
| 60 | 144.24  | 7/2 | -1.065     |
| 65 | 158.924 | 3/2 | 2.014      |

$M_p$ is the proton mass. The modified Coulomb potential due to the interaction between the magnetic moments is given by the expression [2,4]

$$V(r) = \frac{-Ze^2}{1.65\,r}\left[1+\cos\left(\frac{a}{r}\right)\right] \qquad (9)$$

$$a = \frac{\pi e^2}{mc^2} \left( \frac{2m}{M_p} \gamma_N I - m_l \right)$$

where $m_l$ is the magnetic quantum number of the pion. We have found [5] that the solution of the hydrogen atom Schrödinger equation in the constraint condition that the argument of the cosine and sine functions is $a/n^2 a_o$, is

$$E = \frac{-\hbar^2}{2m} [1 + 0.65013266 \cos(a/n^2 a_o) + 0.65013266 \sin(a/n^2) a_o]^2$$

$$\times \frac{[\sqrt{n^2 + \frac{4mca}{144\pi\hbar} \times 1.30026532 (\cos(a/n^2 a_o) - \sin(a/n^2 a_o)N)} - n]^2}{(1.30026532^2) a^2 (\cos(a/n^2 a_o) - \sin(a/n^2 a_o))^2} ; \quad (10)$$

$$N = \frac{1 + 0.65013266 \cos(a/n^2 a_o)}{1 + 0.65013266 [\cos(a/n^2 a_o) + \sin(a/n^2 a_o)]}$$

which for $a/n^2 a_o << 1$ becomes

$$E = \frac{-\alpha^2 mc^2}{2n^2} \left[ 1 - \frac{1.30026532}{3.30026532} \frac{\alpha mca}{n^3 \hbar} \left(1 - \frac{a}{n^2 a_o}\right) \frac{1.65013266}{1.65013266 + 0.65013266 \frac{a}{n^2 a_o}} \right]^2 \quad (10a)$$

where $\alpha = 2 \times 1.65013266/144\pi$. For hydrogen-like atom $\alpha$ is replaced by $\alpha Z$. By using the solution for the hydrogen-like atoms, we can write

$$\langle \psi | [\frac{p^2}{2m} - \frac{Ze^2}{1.65 r} (1 + 0.65 \cos(a/r))] \psi \rangle = E$$

(11)

Further,

$$\Delta E = \langle \Psi | \frac{-(p^2)^2}{8m^3 c^2} \Psi \rangle = \frac{-1}{mc^2} \frac{Z^2 \alpha^2 \hbar^2 c^2}{n^3 a_o^2} \frac{1}{2l+1}$$

$$\frac{-1}{2mc^2} \cdot \frac{Z^4 \alpha^4 m^2 c^4}{4n^4} [1 - \frac{1.30026532}{3.30026532} \frac{Z \alpha mca}{n^3 \hbar} (1 - \frac{a}{n^2 a_o}) \frac{1.65013266}{1.65013266 + 0.65013266 \frac{a}{n^2 a_o}}]^4$$

$$+ \frac{1}{2mc^2} \frac{Z \alpha \hbar c}{n^4 a_o} Z^2 \alpha^2 mc^2 [1 - \frac{1.30026532}{3.30026532} \frac{Z \alpha mca}{n^3 \hbar} (1 - \frac{a}{n^2 a_o}) \frac{1.65013266}{1.65013266 + 0.65013266 \frac{a}{n^2 a_o}}]^2$$

(12)

By using relation

$$E - V(r) = \sqrt{p^2 c^2 + (mc^2)^2}$$

and relations (6) and (10-12) one obtains for the bound state energy $E_b = E - mc^2$, the following

expression

$$E_b = E + \Delta E \quad (13)$$

where $E$ is given by Eq. (10a) and $\Delta E$ is given by Eq. (12).
The results are presented in Table II, $E_b(-)$ when the two magnetic moments substract, and $E_b(+)$ when the two magnetic moments add one another.

Table II

Binding energies of 2p pions for various central charges Z

| Z | $-E_b(-)$, eV | $-E_b(+)$, eV | $-E_b(0)$, eV |
|---|---|---|---|
| 10 | 92930.707320819 | 92930.6752206455 | 92930.974394939 |
| 20 | 372587.68115116 | 372586.78741758 | 372581.51751703 |
| 30 | 841565.39441219 | 841563.6760434 | 841584.48977711 |
| 40 | 1504189.5656131 | 1504189.2593346 | 1504189.6756641 |
| 50 | 236506.2605144 | 2366505.9329577 | 2366512.9901531 |
| 60 | 3436318.6872157 | 3436295.066632 | 3436319.9604135 |
| 65 | 4051933.7664406 | 39360906.914632 | 4051935.2945534 |

The last column presents the binding energy when we neglect the interaction between magnetic moments, that is $a = 0$. For $a \to 0$, relation (13) reduces to the well known expression[1]

$$E_b = mc^2 \left[ \frac{-1}{2} \frac{Z^2 \alpha^2}{n^2} - \frac{Z^4 \alpha^4}{(2l+1)n^3} + \frac{3}{8} \frac{Z^4 \alpha^4}{n^4} \right] \quad (14)$$

where we have used relation $a_o = \hbar/mcZ\alpha$. The number of pions per shell is $(2l+1)$.
From Table II it is observed that the interaction between magnetic moment decreases the absolute value of the binding energy. This decrease is larger when the two magnetic moments, of the nucleus and of the pion, respectively, add one another. In our calculations we have considered $\alpha = 3.30026532/144\pi$.

Note that we have attempted to find the binding energy of the pion by using equation (4) with $E$ given by Eq. (10a). Substituting $\mu = E/c^2$, in this case, Eq. (4) becomes a quarten equation in $E$. We have solved this equation, but numerical results are far away from the acceptable results.

## 3. Conclusions

We have calculated the binding energies of 2p pions for various central charges Z. We have taken into account the interaction between the magnetic moment of the nucleus and the orbital magnetic moment of the pion. The interaction between magnetic moments decreases the absolute value of the binding energy. The derease is larger when the two magnetic moments add one another. The binding energy increases with Z increasing.

## References

1. W. Greiner, *Relativistic Quantum Mechanics*, Springer, 2000.